\documentstyle[12pt]{article}
\input epsf
\def\be{\begin{equation}}
\def\ee{\end{equation}}
\def\l{\label}

\def\eg{{\it e.g.}}
\def\ie{{\it i.e.}}

\def\LEP2{{LEPII}}

\begin{document}
\begin{titlepage}
\vspace*{-1.5cm}
\begin{center}
\null
\vskip-1.5truecm
\rightline{KEK Preprint 97-31}
\vskip1truecm
\null
\vskip-1.5truecm
\rightline{June, 1997}
\vskip1truecm
\vspace{5ex}
{\Large {\bf  Yukawa Unification in Moduli-dominant SUSY Breaking}}\\
\vspace{3ex}
{\bf Shaaban Khalil}$^{a,b}$ and {\bf
Tatsuo
Kobayashi}$^{c}$\\
{\it
\vspace{1ex} a) International Center For Theoretical Physics,\\
 ICTP, Trieste, Italy.}\\
{\it
\vspace{1ex} b) Ain Shams University, Faculty of Science\\
 Department of Mathematics, Cairo, Egypt.}\\
{\it
\vspace{1ex} c)Institute of Particle and Nuclear Studies\\ 
High Energy Accelerator Research Organization\\
Midori-cho, Tanashi, Tokyo 188, Japan.}\\
\vspace{6ex}
{ABSTRACT}  
\end{center}
\begin{quotation}
We study Yukawa unification in string models with 
moduli-dominant SUSY breaking.
This type of SUSY breaking in general leads to non-universal soft masses,
\ie\ soft scalar masses and gaugino masses.
Such non-universality is important for phenomenological aspects 
of Yukawa unification, \ie , successful electroweak breaking, 
SUSY corrections to the bottom mass and the branching ratio 
of $b \rightarrow s \gamma$.
We show three regions in the whole parameter space which 
lead to successful electroweak breaking and allow 
small SUSY corrections to the bottom mass.
For these three regions we investigate the $b \rightarrow s \gamma$ decay 
and mass spectra.
\end{quotation} 
\end{titlepage}
\vfill\eject

\section{Introduction}
\hspace{0.75cm}The origin of fermion masses is one of the most important
problems in particle physics.
Yukawa unification is an attractive idea for fermion masses.
Yukawa unification can be realized by several approaches, 
i.e. grand unified theories (GUTs), superstring theory and 
coupling reduction theories \cite{CR}.
Within the framework of GUT different types of quarks and/or 
leptons are unified into one representation of a GUT group.
On the other hand, the origin of couplings is unique in superstring theory 
including gauge couplings.
Thus superstring theory can realize Yukawa unification especially for 
strong Yukawa couplings without a unified gauge group, although weak Yukawa 
couplings could be realized in terms of higher dimensional 
couplings.
Coupling reduction theories can also realize Yukawa unification without 
unified symmetries, although different types of symmetries might 
be hidden behind coupling reduction theories \cite{CRst}.\\
\vskip -0.3cm
Top-bottom-tau Yukawa unification has been studied in GUTs.
That requires a large value of $\tan \beta \simeq m_t/m_b$, where $m_t$
and $m_b$ are the top and bottom quark masses respectively.
Phenomenological aspects of such models are very different from 
those with small $\tan \beta$.
In particular radiative electroweak symmetry breaking is 
an important issue and actually it has been discussed 
under assumption of universal soft supersymmetry (SUSY) 
breaking parameters \cite{universal,COPW}.
In the large $\tan \beta$ case the mass parameter for the down sector 
Higgs field $H_1$ as well as the up sector Higgs field $H_2$ 
has a large and negative renormalization group equation (RGE) effects 
due to Yukawa couplings.
Hence both of the Higgs mass parameters run from higher energy scale 
to the weak scale 
in a similar way if these masses are universal at the 
Planck scale.
That is unfavorable for successful symmetry breaking.
Thus non-universality such as $m_{H_1}^2 > m_{H_2}^2$ at the 
Planck scale is favorable for successful symmetry breaking 
with Yukawa unification \cite{CW,Nuniver}.
Further non-universality of squark and slepton masses affects 
symmetry breaking \cite{CW,Nuniver} as well as other phenomenological 
aspects \cite{nonU}.
In general such non-universality can  be derived from 
supergravity models.\\
\vskip -0.3cm
Gaugino masses also play a role in phenomenological aspects 
of Yukawa unification.
RGE effects due to gaugino masses generate a significant 
difference between $m_{H_1}$ and $m_{H_2}$ \cite{COPW,CW}.
Thus large gaugino masses are favorable for successful electroweak 
symmetry breaking.
On the other hand, one of SUSY corrections to the bottom mass 
is proportional to the gluino mass \cite{hall}.
Hence a small gluino mass is favorable not to obtain a large 
SUSY correction.
In the universal scenario
such SUSY corrections to the bottom mass are not 
suppressed \cite{COPW,Nuniver}. 
Moreover the
minimal supersymmetric models with electroweak radiative breaking and
universal soft mass terms at the GUT scale is ruled out due to the
constraints from the $b \rightarrow s \gamma$ decay and the condition
$\Omega h^2 <1$ \cite{BOP}. It was shown in Ref.~\cite{BOP,Nuniver} that 
non-universality seems to be mandatory to satisfy these constraints with
small SUSY corrections to the bottom quark.\\
\vskip -0.3cm
SUSY breaking mechanism is a unsolved and important problem 
in SUSY models including superstring models.
However, we can obtain generic formulae for soft SUSY breaking 
parameters assuming which types of fields contribute to 
SUSY breaking \cite{ST-soft}.
Further soft SUSY breaking terms can be parameterized simply by 
the gravitino mass $m_{3/2}$ and goldstino angles in the case 
where the dilaton and moduli fields contribute to SUSY breaking in 
superstring models \cite{BIM,multiT}.
Phenomenological aspects have been discussed in several parts of the 
whole parameter space \cite{stph,stph2}.
In such framework moduli dominant SUSY breaking leads to non-universality  
among soft SUSY breaking terms, which could give 
phenomenologically interesting aspects.
In particular multi-moduli cases can provide strong 
non-universality \cite{multiT}.
Recently a typical string model with multi-moduli dominant SUSY breaking 
is discussed in Ref. \cite{Tmodel}, showing 
several interesting aspects of the small $\tan \beta$ case.\footnote{See also 
Ref. \cite{CDG}.}
Such non-universality between $m_{H_1}^2$ and $m_{H_2}^2$ derived from 
moduli-dominant SUSY breaking is useful also for successful electroweak 
symmetry breaking with Yukawa unification.
Therefore, in this paper we study Yukawa unification of the model 
with non-universal Higgs masses as well as non-universal squark and 
slepton masses within the framework of the moduli-dominant SUSY breaking.\\
\vskip -0.3cm
Further gaugino masses are non-universal in moduli-dominant SUSY breaking 
and their magnitudes are written by a parameter, which is independent 
of the goldstino angle parameterizing non-universality between 
$m_{H_1}^2$ and $m_{H_2}^2$.
Therefore our model is useful not only to show phenomenological 
aspects of moduli-dominant SUSY breaking, but also 
to study what is in general favorable for successful electroweak 
symmetry breaking within Yukawa unification.\\
\vskip -0.3cm
This paper is organized as follows.
In section 2 we review formulae for soft SUSY breaking terms, 
parameterizing them.
We define our model and give its soft terms.
In section 3 we study radiative electroweak symmetry breaking 
in our model with large $\tan \beta$.
In section 4 we consider constraint from SUSY corrections to the 
bottom mass.
Such corrections are usually large.
However, there are three types of parameter regions leading 
to small SUSY corrections.
These correspond to the case with the very light gluino, 
the small $\mu$ case and the small gaugino mass case with 
large squark masses.
Further we discuss constraints from the $b \rightarrow s \gamma$ decay 
in each of these regions.
In section 5 we investigate mass spectra of these three regions, 
in particular masses of the lightest neutralino, chargino and 
Higgs particle.
Section 6 is devoted to conclusions and discussions.
\section{Moduli-dominant SUSY breaking}
\hspace{0.75cm}We assume the string model which has the same massless
matter content 
as the minimal supersymmetric standard model (MSSM), \ie\  
three families of quark doublets $Q_i$, the up-types of quark singlets 
$U_i$, the down-type of quark singlets $D_i$, lepton doublets $L_i$ 
and lepton singlets $E_i$ as well as two Higgs fields.
Here we consider orbifold models with three diagonal moduli fields 
$T_m$ $(m=1,2,3)$ as well as the dilaton field $S$.
We assume dilaton and moduli fields contribute to 
SUSY breaking and the vacuum energy vanishes.
Corresponding $F$-terms are parameterized by the gravitino mass $m_{3/2}$ 
and goldstino angles $\theta$ and $\Theta_m$ as \cite{BIM,multiT} 
\begin{eqnarray}
(K_{S}^S)^{1/2} F^S = \sqrt{3} m_{3/2} \sin \theta, \quad
(K_{T_m}^{T_m})^{1/2} F^{T_m} = \sqrt{3} m_{3/2} \cos \theta \Theta_m,
\end{eqnarray}
where $K^S_S$ and $K_{T_m}^{T_m}$ are K\"ahler metric and 
$\sum_{m=1}^3 \Theta_m^2=1$.
In this case a soft scalar mass of a field with a modular weight $n_i^m$ 
is obtained as \cite{BIM,multiT} 
\begin{eqnarray}
m_i^2 &=& m_{3/2}^2 (1+3 \cos^2 \theta \sum_m n_i^m \Theta_m^2 ).
\label{softm} 
\end{eqnarray}

It is obvious that there appears stronger non-universality among 
soft scalar masses when two fields have nonvanishing elements of 
modular weights corresponding to different moduli fields.
Here we assume modular weights for $H_1$ and $H_2$ as 
\begin{eqnarray}
n_{H_1}=(-1,0,0), \quad n_{H_2}=(0,-1,0).
\end{eqnarray}
Such non-universality becomes strong in the moduli-dominant SUSY breaking 
case, while soft SUSY breaking terms are universal in the 
dilaton dominant SUSY breaking.
We take the limit $\sin \theta \rightarrow 0$.
Further it is favorable that $m^2_{H_1} \geq m^2_{H_2}$.
Thus we take here $\Theta_1=0$.
In this case we always have $m^2_{H_1} \geq m^2_{H_2}$ and 
the other goldstino angles, $\Theta_2$ and $\Theta_3$, can be 
written as $\Theta_2=\sin \theta_1$ and $\Theta_3= \cos \theta_1$.
Using this angle we can write the soft masses of $H_1$ and $H_2$ as 
\begin{eqnarray}
m^2_{H_1}=m_{3/2}^2, \quad m^2_{H_2}=m_{3/2}^2(1-3\sin^2\theta_1).
\label{mh1}
\end{eqnarray}
Thus non-universality is parameterized by $\sin \theta_1$.
The soft mass of $H_2$ could in
principle have a negative mass squared \ie\ $m^2_{H_2}<0$ with a small
magnitude at high energy scale, \ie\ $\sin^2 \theta_1 \geq 1/3$. 
However, in such case one needs fine tuning for other
parameters.
Thus we restrict ourselves to the case with $\sin^2 \theta_1 \leq 1/3$.
As will be seen, we obtain similar results around $\sin^2 \theta \approx
1/3$. Hence we can expect similar results for the case where $\sin^2
\theta_1$ exceeds $1/3$ slightly.\\
\vskip -0.3cm
As will be seen later, RGE effects due to stau masses decrease 
$m^2_{H_1}-m^2_{H_2}$ 
through a large tau Yukawa coupling \cite{COPW,CW}.
Thus small stau masses are favorable for electroweak breaking.
Further the initial condition for squark masses $m_U > m_D$ is also 
favorable for electroweak breaking.
This initial condition is also useful for small SUSY corrections to the 
bottom mass.
Hence we assume the following modular weights for quark and lepton fields,
\be
n_Q=n_U=(-1,0,0),
\ee
\be
n_D=n_L=n_E=(0,-1,0),
\ee
where the family index is omitted, because we assume degeneracy among 
three families.
Under this assumption, fields $Q$ and $U$ have the same soft scalar mass 
as $H_1$, while $D$, $L$ and $E$ have the same scalar mass as $H_2$.\\
\vskip -0.3cm
In the moduli-dominant SUSY breaking case,  gaugino masses and A-terms 
corresponding to 
moduli-independent Yukawa couplings are obtained as \cite{BIM,multiT} 
\begin{eqnarray}
M_a &=& \frac{\sqrt{3} m_{3/2}}{Re S} \sum_m (\frac{b'^m_a}{k_a}-
\delta_{GS}^m) D(T_m) \Theta_m, \\
A_{ijk} &=& -\sqrt{3} m_{3/2} \sum_m (1+n_i^m+n_j^m+n_k^m) \Theta_m ,
\end{eqnarray}
where $k_a$ is a Kac-Moody level,  
$\delta_{GS}^m$ is the Green-Schwarz coefficient \cite{GS} and 
$D(T_m)$ is the moduli-dependent function written by 
the Eisenstein function $\widehat G(T)$
as \footnote{Several kinds of modular functions are shown
in Ref.\cite{CFILQ}.}
\begin{eqnarray}
D(T) = \frac{(T+T^*)}{32\pi^3} \widehat G(T), 
\end{eqnarray}
which takes values as $D(T)= 1.5 \times 10^{-3}$, $2.7 \times 10^{-2}$,
$9.3 \times 10^{-2}$, $0.46$ and $0.66$ for $T=1.2$, $5.0$, $15$, $70$ and
$100$, respectively.
Further $b'^m_a$ are duality anomaly coefficients, which 
depend on modular weights in a model as \cite{DFKZ}
\begin{eqnarray}
b'^m_a = -C(G_a) + \sum_R T(R) (1+2n_a^m),
\end{eqnarray}
where $C(G_a)$ is the casimir of the adjoint representation and
$T(R)$ is the index of the $R$ representation.
In our case we have $b'^m_3=(-6,0,3)$, $b'^m_2=(-5,1,5)$ and 
$b'^m_1=(1,-1,11)$.
Further we take $k_3=k_2=1$ and $k_1=5/3$.
In this case gaugino masses are obtained as
\begin{eqnarray}
M_1= \frac{\sqrt{3} m_{3/2}}{Re S} \left[ (-3/5-\delta_{GS})\sin\theta_1
+(33/5 -\delta_{GS}) \cos\theta_1 \right] D(T),
\nonumber\\
M_2= \frac{\sqrt{3} m_{3/2}}{Re S} \left[ (1-\delta_{GS})\sin\theta_1
+(5 -\delta_{GS}) \cos\theta_1 \right] D(T),\\
\label{gaugino1}
M_3= \frac{\sqrt{3} m_{3/2}}{Re S} \left[ -\delta_{GS}\sin\theta_1
+(3 -\delta_{GS}) \cos\theta_1 \right] D(T).
\nonumber
\end{eqnarray}  
In these equations we have assumed $T_m=T$ and $\delta_{GS}^m=\delta_{GS}$ 
for simplicity and we take $Re S=2$.
In addition the $A$-terms are written as
\begin{eqnarray}
A_t=A_b= -\sqrt{3} m_{3/2} \cos\theta_1, \quad 
A_\tau= -\sqrt{3} m_{3/2} (-\sin \theta_1 + \cos\theta_1).
\label{trilinear}
\end{eqnarray}

The overall magnitude of the gaugino masses is dominated by $D(T)$ and 
their ratios depend on $\delta_{GS}$.
We have $M_3(M_Z) > M_2(M_Z) > M_1(M_Z)$ in most of 
the parameter space of $\delta_{GS}$. It is important to notice that
the gaugino masses are parameterized by $D(T)$ and $\delta_{GS}$ 
independent of
the other soft SUSY breaking terms parameterized by $\theta_1$ 
as well as $m_{3/2}$. 
As shall be shown later, the gaugino masses are very important for 
phenomenological aspects of Yukawa unification, 
electroweak symmetry breaking and SUSY corrections to the bottom mass.
A large gaugino mass is favorable for successful electroweak symmetry 
breaking.
On the other hand, a large gluino mass leads to a large SUSY correction 
to the bottom mass.
Thus this parameterization (\ref{gaugino1}) as well as eq.(\ref{mh1}) is 
quite interesting not only to study phenomenology of moduli-dominant 
SUSY breaking, but also to investigate which types of spectra and 
non-universality are favorable for generic models with Yukawa unification, 
in particular from viewpoints of successful electroweak symmetry breaking,
small SUSY corrections to the bottom mass and the experimental bounds 
of the $b \rightarrow s \gamma$ decay.
Moreover, for $\delta_{GS} \simeq 3/(1+\tan \theta_1) $, we obtain 
very light gluino. 
The possibility of having the light gluino of order $1-4$ GeV
is not excluded experimentally. In the models with the universal gaugino mass
there are difficulties to get light gluino with proper 
radiative breaking of the electroweak symmetry satisfying 
experimental constraints. We will
show that in our case we can have a part of the parameter space which
leads to a very light gluino and satisfies all other constraints. 
This possibility of the light gluino is a quite interesting solution 
for controlling the SUSY correction $\delta m_b$.\\ 
\vskip -0.3cm
For the $B$-term its magnitude depends on a way to generate a natural 
$\mu$-term.
Therefore here we take $\mu$ and $B$ as free parameters and 
we fix them requiring successful electroweak symmetry breaking.
\section{Electroweak symmetry breaking in Yukawa unification}
\hspace{0.75cm}We assume the equality of the top, bottom and tau
Yukawa couplings at the string scale, {\it i.e.},
$\lambda_t = \lambda_b =\lambda_{\tau}=\lambda_G$.
The RGEs of these couplings are obtained
as follow:
\begin{eqnarray}
\frac {d Y_t}{d t}&=& Y_t \left( \frac{16}{3} \tilde{\alpha_3} + 3 
\tilde{\alpha_2} + \frac{13}{9} \tilde{\alpha_1}-6 Y_t -Y_b \right ),
\\
\frac {d Y_b}{d t}&=& Y_b \left( \frac{16}{3} \tilde{\alpha_3} + 3
\tilde{\alpha_2} + \frac{7}{9} \tilde{\alpha_1} -  Y_t -6 Y_b
-Y_{\tau}\right ),
\l{yb}
\\
\frac {d Y_{\tau}}{d t}&=&Y_{\tau} \left( 3 \tilde{\alpha_2} + 3 
\tilde{\alpha_1} - 3 Y_b -4Y_{\tau} \right ),
\end{eqnarray}
where $Y_i = \frac{\lambda_i^2}{(4 \pi)^2}$, and $i=t,b,\tau$.
The pole masses of the top quark $M_t=175$ GeV and the tau lepton
$m_{\tau}=1.78$ GeV are used to determine the common value of the Yukawa
coupling $\lambda_G$ as well as the corresponding values of $\tan \beta$.
Hence we can estimate the tree level mass of the $b$-quark, \ie\  
without SUSY corrections, $m_b=\lambda_b v \cos \beta$, where 
$v=\sqrt{\langle H_1^0
\rangle ^2 + \langle H_2^0 \rangle ^2}=174$ GeV. \\
\vskip -0.3cm
        The fermion masses at $M_Z$-scale are related to their Yukawa
couplings and we have 
\begin{equation}
\frac{m_t^2(M_Z)}{Y_t(M_Z)} + \frac{m_{\tau}^2(M_Z)}{Y_{\tau}(M_Z)} =
(4\pi)^2 v^2, 
\end{equation}
which we solve for single unknown $\lambda_G$, then we find 
$\lambda_G \simeq 0.32 $. This value leads to $\tan \beta \simeq 50$ and
$m_b(M_Z) \simeq 3.3$ GeV. The experimental value of the bottom quark mass
still has some uncertainty. For instance, the analysis of the
$\Upsilon$ system using QCD sum rules 
\cite{jamin} gives $m_b(m_b) = 4.13 \pm 0.06$ GeV
corresponding to $m_b(M_Z) = 2.83 \pm 0.10$ GeV. 
On the other hand, the last lattice result
shows \cite{gimenez} $m_b(m_b)=4.15 \pm 0.20$ GeV and 
$m_b(M_Z) = 2.84 \pm 0.21$
GeV. Moreover, as mentioned in Ref.\cite{santamaria}, the DELPHI
collaboration \cite{fuster:97} extracted $m_b(M_Z)$ to be 
\begin{equation}
m_b(M_Z) =  2.85 \pm 
0.22~({\rm stat})\pm 0.20~({\rm theo}) 
\pm 0.36~({\rm fragmentation})~GeV.  
\end{equation}
The lower bound of $m_b(M_Z)$ in this case is around 2.15 GeV. This
could allow for a large negative SUSY correction to $m_b$, 
$\delta m_b \sim 35 \%$.
However, here
we will be conservative and we will consider the lower bound of $m_b(M_Z)$ 
as 2.63 GeV. Hence SUSY corrections have to be negative and of order
$\delta m_b \leq 20 \%$.
These SUSY corrections will be discussed in the next section.\\  
\vskip -0.3cm
Let us now study electroweak symmetry breaking. The Higgs potential is
written as 
\begin{eqnarray}
V(H_1,H_2) & = & \frac{1}{2} g^2(H_1^*\frac{\tau^a}{2} H_1 +
H_2\frac{\tau^a}{2}H_2)^2 +
\frac{1}{2} g'^2(\mid H_2\mid^2 - \mid H_1\mid^2)^2
        \nonumber\\
            & + &m_1^2 \mid H_1\mid^2 + m_2^2\mid H_2\mid^2 - m_3^2(H_1
H_2 + h.c),
\end{eqnarray}
where
\begin{equation}
m^2_i = m^2_{H_i} + \mu^2 , \hspace{0.5cm} i=1,2 \hspace{2cm}  
m_3^2 = - B \mu.
\l{m12}
\end{equation}
We take $\mu$ and $B$ as free parameters and these are fixed by potential 
minimization conditions.
In the large $\tan \beta$ case the above Higgs potential has 
two characteristic features. It
follows from the minimization conditions that
\be
m_2^2 \simeq -{M_Z^2 \over 2},
\label{cons1}
\ee
\be
m_3^2 \simeq {M_A^2 \over \tan^2 \beta}\sim 0,
\label{cons2}
\ee
with
\be
M_A^2 \simeq m_1^2 +m_2^2 >0.
\label{cons3}
\ee
A combination of eqs.(\ref{cons1}) and (\ref{cons3}) gives the following 
constraint on the low energy parameters
\be
m_1^2 -m_2^2 > M_Z^2
\label{cons4}
\ee
i.e $m_{H_1}^2 -m_{H_2}^2 > M_Z^2$. 
In order to have electroweak breaking in the large $\tan \beta$
case, the difference between the masses of the two Higgs fields 
should satisfy the above inequality. 
\\
\vskip -0.3cm
In general case of the non-universal soft SUSY breaking terms, we find
that the mass difference between $m_{H_1}^2$ and $m_{H_2}^2$ 
at the weak scale is given
by
\begin{equation}
m_{H_1}^2 -m_{H_2}^2 = \Delta m^2 +\alpha_{ab} M_a M_b + \beta_k m_k^2
+\gamma_i A_i^2 + \lambda_{ia} A_i M_a
\label{dif}
\end{equation}
where $ \Delta m^2$ is given by $ \Delta m^2=
m_{H_1}^2-
m_{H_2}^2 $ at the string scale. In our case we have 
$ \Delta m^2=3 m_{3/2}^2\sin^2 \theta_1$. Here $m_k$ refer to $m_Q$,
$m_U$, $m_D$, $m_L$ and $m_E$ at
the string scale and 
$A_i$ are the trilinear couplings $A_t$, $A_b$ and $A_{\tau}$.\\
\vskip -0.3cm
        In the case with universal soft SUSY breaking parameters, 
\eg\ the universal scalar mass $m_0$ and the universal gaugino mass 
$M_{1/2}$, \cite{COPW} we find
\begin{equation}
\gamma_i \simeq \lambda_{ij} \simeq 0,
\end{equation}
\begin{equation}
\alpha_{ab}M_{ab} =\alpha M_{1/2}, \quad  \beta_k m^2_k=\beta m_0
\end{equation}
with $0.1 \leq \alpha \leq 0.2$ and $\beta \simeq -0.2$.
The tau Yukawa coupling contributes dominantly to the third term of R.H.S. 
in eq.(\ref{dif}).
Constraints such as $\mu \simeq (1.5-1.7)M_{1/2}$ and $
M_{1/2}>m_0$ are necessary for successful electroweak symmetry breaking.
A lower bound on the universal gaugino mass 
$M_{1/2}$ of order 300 GeV is obtained to make $m_{H_2}^2$ smaller than
$m_{H_1}^2$.
Thus large gaugino masses as well as a small stau mass 
are favorable for successful electroweak symmetry breaking.
This statement is still true in generic models.
On the top of that, the non-universality $\Delta m^2$ 
as well as non-universality $m^2_U-m^2_D$ is favorable 
for electroweak symmetry breaking.\\
\vskip -0.3cm
In our model the difference $m_{H_1}^2-m_{H_2}^2$ is written in terms of
$m_{3/2}$, $\sin \theta_1$, $D(T)$ and $\delta_{GS}$.
When we fix values of $D(T)$ and $\delta_{GS}$, this difference is 
written as 
\begin{equation}
m_{H_1}^2 -m_{H_2}^2 \simeq (a \sin^2 \theta_1 +b)m_{3/2}^2.
\label{dif2}
\end{equation}
For example in the case with $D(T)=0.23$ and $\delta_{GS}=-5$  
we find $a \simeq 3.87$ and $b \simeq 0.4$ in eq.(\ref{dif2}). 
It is clear that in this case
we have $m_{H_1}^2 -m_{H_2}^2 > M_Z^2$ 
for all values of $\theta_1 \in [0,0.6]$ and 
the constraint (\ref{cons4}) leads to $m_{3/2} \geq 150$ GeV.
Thus the gaugino mass corresponding to $D(T) \geq O(0.1)$  is 
large enough to lead to successful electroweak symmetry breaking 
for all the range of $\theta_1$ even without non-universality $\Delta m^2$.
Further this case does not require very heavy gravitino mass.\\
\vskip -0.3cm
On the other hand, in the case with 
$D(T)=0.027$ and $\delta_{GS}=-5$, 
we find $ a \simeq 3.5$, $ b \simeq -0.16$ in eq.(\ref{dif2}).
Now to require $m_{H_1}^2
-m_{H_2}^2 > M_Z^2$ we obtain constraints in both $m_{3/2}$ and
$\theta_1$. Moreover we
also require that the mass of the psuedoscalar Higgs $m_A$ satisfies the
experimental constraint {\it i.e }\ $m_A \geq 40$ GeV. Combining this
constraint with the above ones we find that $m_{3/2} \geq 300$ GeV and
$\theta_1 > 0.2$ rad. Therefore we conclude that in our model with 
non-universality between $m_{H_1}^2$ and $m_{H_2}^2$ the scalar
masses need no longer be smaller than gaugino masses. Further, as we will
show in the next section, the large hierarchy $(m_i >> M_a)$ is favoured
for obtaining small SUSY correction to the bottom quark mass. This shows
that the value of $D(T)$ plays an important role in
studying the Yukawa unification scenario.
\section{ SUSY corrections to $m_b$ and the $b\rightarrow s
\gamma$ decay rate}
\hspace{0.75cm}
In this section we calculate SUSY corrections to the bottom mass 
in the model with successful electroweak symmetry breaking. 
We are interested in finding regions of the parameter space
$m_{3/2}$, $\theta_1$, $T$ and $\delta_{GS}$ which allow small SUSY
corrections to $m_b$ and at the same time have proper electroweak breaking.\\
\vskip -0.3cm
The bottom mass receives SUSY corrections as 
 $m_b=\lambda_b v_1 (1+\delta m_b)$.
Here dominant contributions to $\delta m_b$ are due to the
sbottom-gluino and stop-chargino loops, given
in Ref.~\cite{hall} 
\begin{eqnarray}
\delta m_b&=&\frac{2 \alpha_3}{3\pi} M_{\tilde{g}} \mu \tan \beta\
 I(m_{\tilde{b}1}^2,m_{\tilde{b}2}^2, M_{\tilde{g}}^2)
\nonumber
\\
        &+& Y_t A_t \tan \beta\ I(m_{\tilde{t}1}^2,m_{\tilde{t}2}^2,
M_{\tilde{g}}^2), 
\label{bcor}
\end{eqnarray} 
where $M_{\tilde{g}}$, $m_{\tilde{b}i}$ and $m_{\tilde{t}i}$ are the
gluino, sbottom and stop eigenstate masses respectively. The integral
function $I(a,b,c)$ is given by
\begin{equation}
I(a,b,c)= \frac{ab\ln(a/b) +bc \ln(b/c) + ac \ln (c/a)}{(a-b)(b-c)(a-c)}.
\end{equation} 
The function $I(a,b,c)$ is of order $1/m_{max}^2$ where $m_{max}$
is the largest mass running in the corresponding loop. 
The first term of R.H.S. in eq.(\ref{bcor}) is rather dominant.
For small SUSY corrections models should satisfy at least 
one of the following conditions, 
$M_{\tilde{g}} \ll m_{\tilde{q}}$ or $\mu \ll m_{\tilde{q}}$ 
where $ m_{\tilde{q}}$ represent the heaviest mass of the 
third generation squark 
eigenstates. By scanning the parameter space of this model, we find that
there are three different regions which lead to $\delta m_b \leq 20 \%$.
These regions correspond to very light gluino of 1--4GeV, small $\mu$  
and small gaugino mass of $O(100)$GeV with large
sparticle. In all these regions large non-universality
between $m_{H_1}^2$ and $m_{H_2}^2$ is favorable.
Thus $\theta_1$ runs from 0.3-0.6 rad. \ie\ around $\sin^2\theta = 1/3$ to
give maximum non-universality between $m_{H_1}^2$ and $m_{H_2}^2$.
\subsection{Light gluino scenario}
\hspace{0.75cm}As mentioned above, at $\delta_{GS} \simeq 3/(\tan\theta_1 
+1)$ the value of $M_3$ is
very close to zero while $M_1$ and $M_2$ are of order $m_{3/2}D(T)$.
In this case we have strong non-universality between $M_3$ and other 
gaugino masses. For such value of $\delta_{GS}$ the gluino mass $M_{\tilde
g}$ is of order few GeV, then the SUSY correction $\delta m_b$ becomes
small.  For other values of $\delta_{GS}$, similarly small values of $M_3$
can  be obtained if $D(T)$ is of order $10^{-3}$, \ie\ $T\simeq O(1)$.
However, in this case $M_1$ and $M_2$ are also suppressed and then  
it is difficult to satisfy the \LEP2 lower
bound on the chargino mass. Hence we will not consider such case, but 
we will concentrate the case with $\delta_{GS}$ very close to 
$3/(\tan\theta_1+1)$, which leads to $M_{\tilde g} < 10$GeV. 
In this case the value $D(T)>0.1$ is required to have 
proper electroweak breaking. The corresponding
values of $\delta m_b$ for this region of the parameter space is
found to be smaller than $10\%$.\\
\vskip -0.3cm
        Now, we focus on the constraints from $b \rightarrow s
\gamma$ decay \cite{masiero}. 
In SUSY models, there are additional contributions to the
decay besides the SM one. For large $\tan\beta$, the dominant
supersymmetric contribution comes from the
charged Higgs and chargino exchanges. As explained in Ref.\cite{stph2}
the chargino exchange contribution is enhanced for large $\tan \beta$, and
it becomes sizable. The chargino contribution  
could give rise to a substantial destructive interference with SM and   
Higgs amplitudes, depending on the sign of $\mu$. We follow the
procedure and the notation convention of the sign of $\mu$ in the chargino
mass matrix suggested in Ref.\cite{stph2}.\\
\vskip -0.3cm
Before discussing the constraints of $b \rightarrow s \gamma$ 
on the parameter space, it is worthwhile to show the correlation 
between the masses
of the lightest chargino and charged Higgs. This is   
helpful in understanding our estimation to the branching ratio (BR) of $b
\rightarrow s \gamma$. In Fig.1, we plot the charged Higgs mass 
versus the lightest chargino mass for $\mu
<0$~\footnote{According to our convention, the chargino is heavier in   
the case of $\mu >0$.}.
\vskip 0.3cm
\begin{center}
\input charg.tex
\end{center}
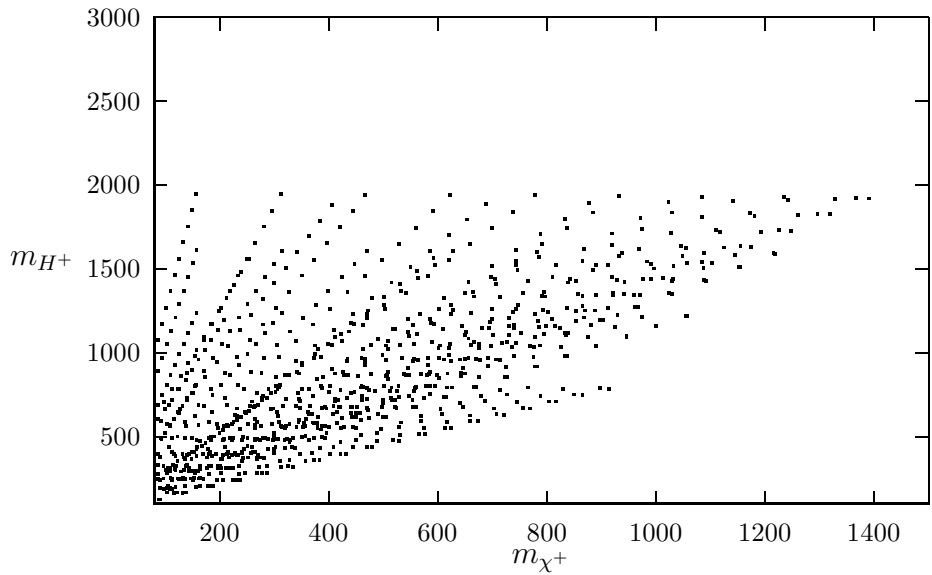
Figure 1: The mass of charged Higgs versus the mass
of lightest chargino, and $\mu <0$.\\
 
We notice from this figure that the charged Higgs tends to be heavier than
the lightest chargino. This feature is very special for this region while 
in most of the parameter space the charged Higgs is lighter. 
In Figs.2 we show the values of the $BR(b \rightarrow s \gamma)$
corresponding to the values of $D(T)$
in the light gluino region of the parameter space we have
determined above for $\mu <0$.
In this case we find 
the chargino contribution gives destructive
interference with the charged Higgs amplitude. For $\mu >0$ the chargino
contribution is additive to charged Higgs and the SM ones, so that
the total branching ratio becomes larger than the experimental limit
$4\times 10^{-4}$.
\vskip 0.3cm
\begin{center}
\input lgbs2.tex
\end{center}
\hspace{2.5cm}Figure 2: $BR(b \rightarrow s \gamma)$ as a function of
$D(T)$, and $\mu <0$.\\
\vskip 0.3cm
It is remarkable that the experimental limits $1 \times
10^{-4} \leq BR(b\rightarrow s \gamma) \leq 4\times 10^{-4}$ impose an
upper bound on the value of $D(T)$, $D(T)\leq 0.45$. 
The value of $\mu$ is decreasing as $D(T)$ increases.
Such behavior is shown in Fig.3 
for $m_{3/2}= 1$ TeV, $\sin^2\theta_1 \simeq 1/3$ and $\delta_{GS}
\simeq 2.1$. 
\vskip 0.3cm
\begin{center}
\input mudt.tex
\end{center}
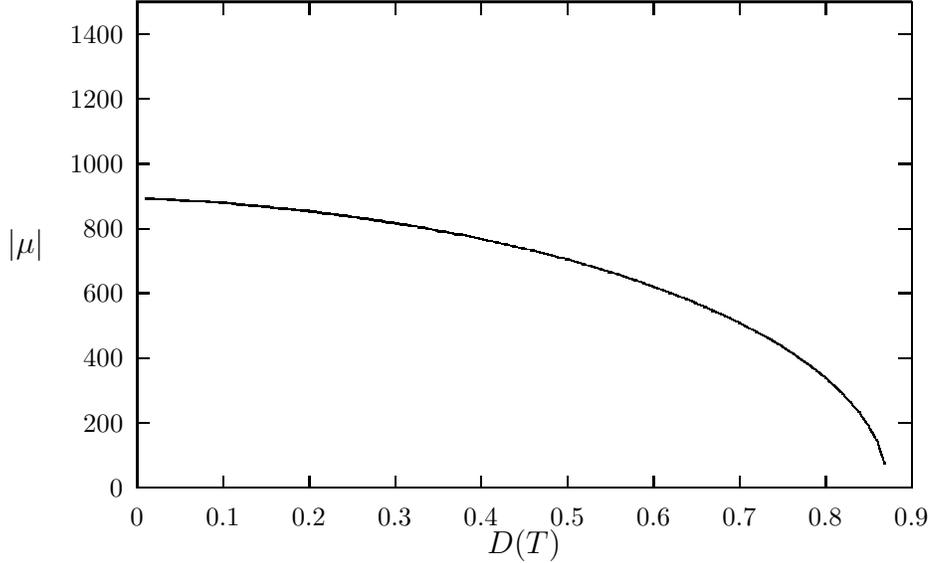
\hspace{3.5cm}Figure 3: The value of $\vert \mu \vert$ as a function of $D(T)$.\\
\vskip 0.3cm
For $D(T) \geq 0.45$ we find that the angle $\phi$, where
\be
\tan 2\phi= 2\sqrt{2} m_W \frac{-\mu \sin \beta +M_2 \cos\beta}{M_2^2
-\mu^2 - 2 M_W^2 \cos 2\beta} ,
\ee
changes its sign.
Note that $M_2$ is increasing as $D(T)$ increases.
This angle determines one of the unitary matrices
that diagonalise the chargino mass matrix and it has contribution to the
chargino amplitude~\cite{stph2}. Then the sign of the chargino
contribution is also changed and it becomes additive. Hence we find
that the value of the $BR(b\rightarrow s \gamma)$ becomes larger than the
experimental upper bound for $D(T)\geq 0.45$ as Fig.2 shows. The same is
happening for $\mu>0$, and due to this sign changing the $BR(b\rightarrow
s \gamma)$ jumps from values larger than the upper bound $ 4 \times
10^{-4}$ to values lower than the lower bound $1 \times 10^{-4}$. Actually
there is the region which satisfies the $b\rightarrow s\gamma$ 
constraint for $\mu >0$, but it is very narrow. 
Moreover for each value of $0.1\leq D(T) \leq
0.45$ it is observed from Fig.2 that the experimental limits impose
another constraints on $m_{3/2}$. For instance, for $D(T)=0.1$, the
gravitino mass should be larger than 600 GeV.  
\subsection{Small $\mu$ scenario}
\hspace{0.75cm}There appears a parameter region with small $\mu$
enough to lead to $\delta m_b\leq 20\%$. In our model such small $\mu$ is 
obtained as $m_{3/2} >200$GeV, large $D(T)$ such as $[0.6,1]$ and 
$\delta_{GS}\simeq 3/(1+\tan \theta_1)$, say 1.5--1.9.\footnote{
Further a value of $\delta_{GS}<1.5$ makes the gluino very heavy 
increasing $\delta m_b$.}
Here the value $D(T)=0.6$ corresponds to $T \sim 90$.
For this region Fig.4 shows values of $\mu$
against the gluino mass $M_{\tilde g}$.
\vskip 0.3cm
\begin{center}
\input mumg.tex
\end{center}
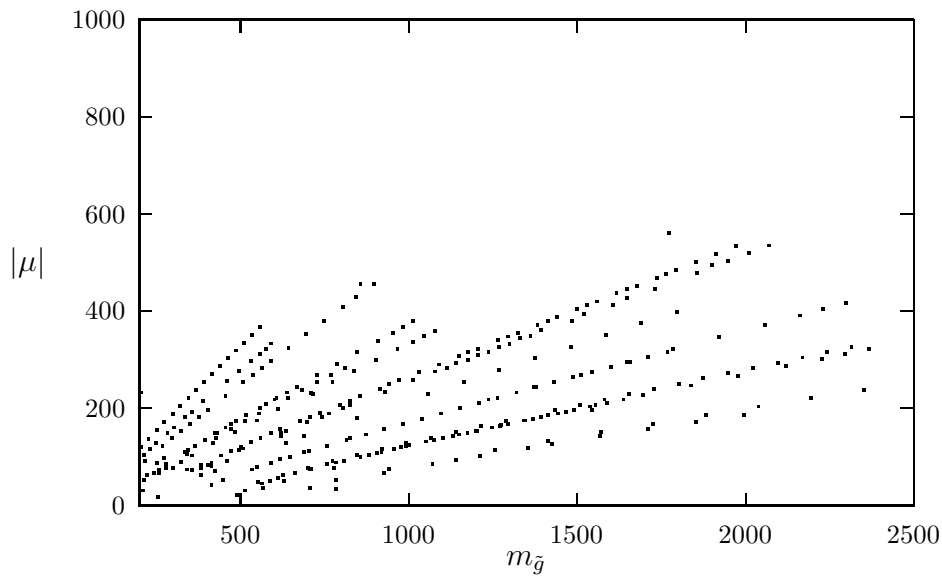
\hspace{3.5cm}Figure 4: The value of $\vert \mu \vert $ versus the gluino
mass.\\
\vskip 0.3cm
This figure shows the ratio of $\vert \mu \vert /M_{\tilde g}$ is less
than $1/5$ for  
very heavy gluino while $\mu$ is about half of $M_{\tilde g}$ 
for smaller gluino mass.
For the latter case suppression of $\delta m_b$ is due to smallness of 
the product $\mu M_3$ compared with $\tilde{m}^2_b$. 
Further Fig.5 shows $BR(b
\rightarrow s \gamma)$ for this region with $\mu <0$.
\vskip 0.3cm
\begin{center}
\input smubs.tex
\end{center}
\hspace{1.5cm}Figure 5: $BR(b \rightarrow s \gamma)$ as a function of
$m_{\chi^+}$, and $\mu <0$.\\
\vskip 0.3cm
As a result, the $b \rightarrow s \gamma$ decay requires no more 
constraints on this parameter region.
\subsection{Small gaugino mass scenario with large sparticle}
\hspace{0.75cm}An alternative scenario for making $\delta m_b \leq 20\%$
is to have small gaugino masses of $O(100)$GeV with large sparticle of 
order TeV. For
instance, we find that for $m_{3/2} \simeq 2$ TeV, and $D(T)\simeq
0.04-0.2$ we can obtain small SUSY correction to $m_b$.\\
 
        The computation of the $b \rightarrow s \gamma$ shows that
the value of the $ BR(b \rightarrow s \gamma)$ is decreased by
increasing the value of $D(T)$.
In this region we find that the values of the branching ration are within
the experimental limits and no further constraints are obtained. 
\section{SUSY spectrum in Yukawa unification}
\hspace{0.75cm}In this section, we investigate the
SUSY spectrum in the three regions of the parameter space which lead to
small SUSY correction to $m_b$.\\
\vskip -0.3cm
In the region corresponds to very light gluino, the lightest neutralino is
found to be $O(100)$ GeV. It is not the lightest supersymmetric
particle (LSP), because obviously the gluino is much lighter.
Hence the lightest neutralino is unlike a dark matter candidate 
in this region. 
However, in string models there are many other candidates for 
dark matter like for example the axion or other singlet fields. 
The lightest chargino in this
region is found to be of order 100 GeV and we have imposed the
experimental constraint $m_{\chi^+} > 84$ GeV in determining this region.
The mass of the lightest Higgs in this region lies between 80-125 GeV.\\
\vskip -0.3cm
On the other hand, in the small $\mu$ region, we find 
that the lightest neutralino is
the LSP and its mass is given in Fig.6 as a function of the gravitino mass
\vskip 0.3cm
\begin{center}
\input smuegn.tex
\end{center}
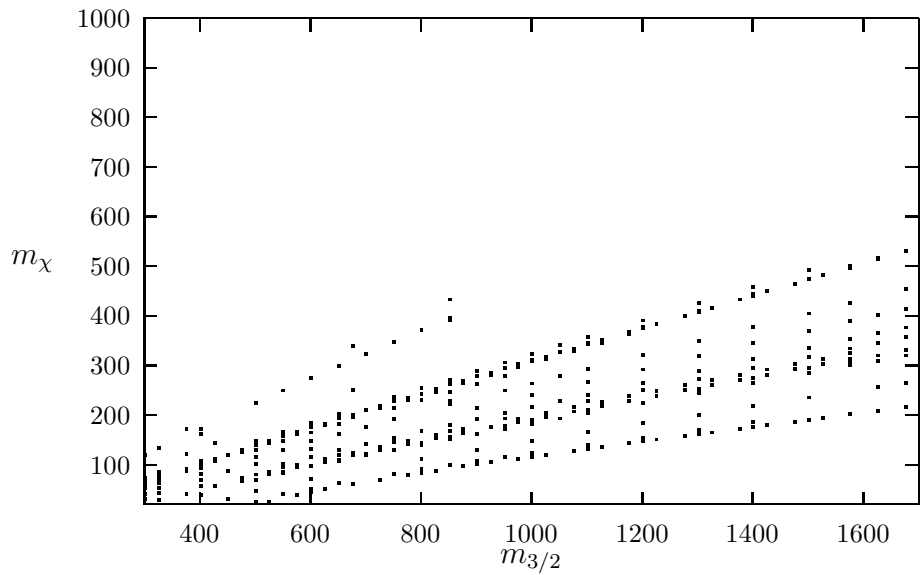
\hspace{1.5cm}Figure 6: The mass of the LSP in the region of small $\mu$ 
as a function of $m_{3/2}$.\\
\vskip 0.3cm
In this region a small mass of $m_\chi$ is allowed.
Thus this LSP becomes a dark matter candidate.
Further we have $M_2 > \mu$, hence the lightest chargino mass is of
order $\mu$. The lightest Higgs mass $m_h$ in this region satisfies
$100$ GeV $<m_h<130$ GeV. \\
\vskip -0.3cm
Finally for the region with small gaugino masses of $O(100)$GeV 
and the scalar masses of
order TeV we find that the lightest neutralino is the LSP and its mass and
the mass of the lightest chargino are of $O(100)$ GeV. Further the
mass of the lightest Higgs is about 125 GeV.

\section{Conclusions}

We have studied Yukawa unification within the framework of 
superstring models with moduli-dominant SUSY breaking.
Large non-universality between $m_{H_1}$ and $m_{H_2}$ as well as 
non-universality $m^2_U-m^2_D$ and small stau masses is 
favorable for successful electroweak symmetry breaking.
Large gaugino masses $M_a$ are also suitable.
On the other hand, SUSY corrections to the bottom mass are 
in general large.
To obtain sufficiently small SUSY corrections gives strong constraints on 
allowed parameter regions.
There appear three types of regions leading to small SUSY corrections.
These correspond to the very light gluino scenario with the mass 
of order $1-4$GeV, the small $\mu$ scenario and the small gaugino mass
scenario 
with heavy squark masses.
In the very light gluino scenario the $b \rightarrow s \gamma$ decay 
constrains gaugino masses as $D(T) \leq 0.45$, \ie\  
$M_a/m_{3/2} \leq  O(0.5)$ for $a=1,2$.
For the other two scenarios, we have no further 
constraint from $BS(b \rightarrow s \gamma)$.
There also appears the region where the lightest neutralino is light 
enough for a dark matter candidate in the small $\mu$ scenario.
This scenario requires a large value of $T \sim O(100)$ 
corresponding to $D(T) > 0.6$.
Such large value for $T$ might not be natural.
The value of $T$ is determined by a nonperturbative mechanism.
It is not clear that such large value could be realized by any mechanism.
However, in the whole parameter space one could find out regions 
with small $\sin \theta$, 
but not exactly $\sin \theta =0$ which lead to spectra similar to 
the case with $D(T) > 0.6$ and $\sin \theta =0$.\\
\vskip -0.3cm
Gauge symmetry breaking can lead to another source of non-universality 
of soft scalar masses, \ie\ $D$-term contributions \cite{Dterm1}.
Such non-universality could also lead to interesting aspects to be 
studied \cite{Dterm2}.\\

\noindent{\Large\bf Acknowledgements}
\vskip0.5truecm
 The authors are grateful to Y.Kawamura, A.Santamaria, Q.Shafi, C.Wagner 
and Y.Yamagishi for useful discussions. S.K would like to acknowledge 
the hospitality of ICTP.

\end{document}